\documentclass[aps,pra,superscriptaddress,amsmath,amssymb,preprintnumbers,floatfix,showpacs,12pt]{revtex4}
\usepackage{amssymb}
\usepackage{epsfig}

\begin{document}
\title{ Relationship between the linear entropy, the von Neumann entropy
and the atomic Wehrl entropy for the Jaynes-Cummings model }
\author{Faisal A. A. El-Orany }
\email{el_orany@yahoo.com} \affiliation{ Department of Mathematics
and Computer Science, Faculty of Science, Suez Canal University,
Ismailia, Egypt; }

\begin{abstract}
 The linear entropy, the von Neumann
entropy and the atomic Wehrl entropy are frequently used to quantify
 entanglement in the quantum systems. These  relations provide
typical information on
 the entanglement in the Jaynes-Cummings model (JCM).
In this Letter,   we explain the origin of this analytically  and
derive a closed form for the atomic Wehrl entropy. Moreover, we
show  that it is more convenient to use the Bloch sphere radius
 for quantifying  entanglement in the JCM instead
of these entropic relations.

\end{abstract}
  \maketitle

Quantum entanglement is an extensively-studied topic in the recent
years in the advent of growing realizations and applications in
the quantum information processing such as  quantum computing
\cite{prin}, teleportation \cite{tel1}, cryptographic \cite{cry1},
dense coding \cite{dens} and entanglement swapping \cite{swap}.
Entanglement gives new insights for understanding many physical
phenomena including super-radiance \cite{lamb}, superconductivity
\cite{vedr}, disordered systems \cite{duw}, etc. Various types of
experiments have been dealt with the entanglement, e.g.
long-distance entanglement \cite{peng},
 ion-photon entanglement \cite{volz}, many photons entanglement \cite{zhao},
  etc. For recent review reader can consult \cite{rysz}.

The notion of entropy, originating from thermodynamics, has been
reconsidered in the context of classical information theory
\cite{shann} and quantum information theory \cite{neum}. There are
several definitions for entropy. For instance, the von Neumann
entropy \cite{neum}, the relative entropy \cite{vdr}, the
generalized entropy \cite{basti}, the Renyi entropy \cite{reny},
the linear entropy, and the Wehrl entropy \cite{wehrl}. The Wehrl
entropy
 has been successfully applied in the description of
different properties of the quantum optical fields such as
phase-space uncertainty \cite{mira1,{mira2}}, quantum interference
\cite{mira2}, decoherence \cite{deco,{orl}},
 a measure of both
noise (phase-space uncertainty) and phase randomization
\cite{mira3}, etc. Also it  has been applied to the evolution of
the radiation field with
 the Kerr-like medium  \cite{jex}
as well as  with the two-level atom \cite{orl}. Quite recently,
 the Wehrl entropy has been used in quantifying the
entanglement of pure states of $N\times N$ bipartite quantum
systems \cite{flor}. Moreover, the concept of the atomic Wehrl
entropy  has been developed  \cite{karol} and applied to the
atom-field interaction  \cite{obad}.

One of the elementary  models in quantum optics, which describes
the interaction between the radiation field and the matter, is the
Jaynes-Cummings model (JCM) \cite{jay1}. The JCM  is a rich source
for the nonclassical effects, e.g., \cite{bruce}. Most importantly
 the JCM has been experimentally implemented by several means,
e.g. one-atom mazer \cite{remp}, the NMR refocusing \cite{meu}, a
Rydberg atom in a superconducting cavity \cite{sup} and  the
trapped ion \cite{vogel}. The JCM is a subject of continuous
studies. Previous investigations for the JCM  have been shown that
the von Neumann entropy \cite{knig}, the linear entropy, \cite{fa}
and the atomic Wehrl entropy \cite{obad} provide typical dynamical
behaviors (N.B. The references given here are just examples,
however, there are a large number  of articles dealt this issue).
In all these studies  the attention is focused on the numerical
investigations only and hence there was no clear answer to the
question: Why these quantities give  typical dynamical behaviors?
In this Letter we answer this question using straightforward
calculations. Moreover, we derive a closed form for the atomic
Wehrl entropy. Additionally, we show that for quantifying
entanglement in the JCM one should use the Bloch sphere radius (,
i.e., the length of the Bloch sphere vector \cite{ekert} )
 instead of  these three quantities.
    These are interesting results and  will be useful  for the scientific
community.

We start the investigation by describing the system under
consideration and giving the basic relations and equations, which
will be frequently used in this Letter. The simplest form of the
JCM is the two-level atom interacting with the single  cavity
mode. In the rotating wave approximation and dipole approximation
the Hamiltonian controlling this system is:
\begin{eqnarray}
\begin{array}{lr}
\frac{\hat{H}}{\hbar}=\hat{H}_0+\hat{H}_i\\
\\
 \hat{H}_0 =
\omega_{0}\hat{a}^{\dagger}\hat{a}+
\frac{1}{2}\omega_{a}\hat{\sigma}_{z},\quad \hat{H}_i=
\lambda(\hat{a}\hat{\sigma}_{+} + \hat{a}^{\dagger
}\hat{\sigma}_{-}),
 \label{6}
 \end{array}
\end{eqnarray}
where $\hat{H}_0\quad (\hat{H}_i)$ is the free (interaction) part,
$\hat{\sigma}_{\pm}$ and $\hat{\sigma}_{z}$ are the Pauli spin
operators; $\omega_{0}$ and $\omega_{a}$ are the frequencies of
the cavity mode and the atomic transition, respectively, $\hat{a}
\quad (\hat{a}^{\dagger})$ is the annihilation (creation) of the
cavity mode, and $\lambda$ is the atom-field coupling constant. We
assume that the field and the atom are initially prepared in the
coherent state $|\alpha\rangle$ and the excited atomic state
$|e\rangle$, respectively, and $\omega_{0}=\omega_{a}$ (, i.e. the
resonance case). Under these conditions the dynamical wave
function
 of the system in the interaction picture can be expressed as:
\begin{equation}
|\Psi (T)\rangle= \sum\limits_{n=0}^{\infty} C_n \left[
\cos(T\sqrt{n+1})|e,n\rangle
-i\sin(T\sqrt{n+1})|g,n+1\rangle\right],
 \label{a8a}
 \end{equation}
 where
 \begin{equation}
C_n=\frac{\alpha^n}{\sqrt{n!}} \exp(-\frac{1}{2}\alpha^2),\qquad
\alpha=|\alpha|\exp(i\vartheta), \qquad T=t\lambda, \label{1}
 \end{equation}
 $|g\rangle$ denotes the atomic ground  state and $\vartheta$ is a phase.
 Information about the bipartite ( i.e., atom and field)  is
involved in the wavefunction (\ref{a8a}) or in the total density
matrix $\hat{\rho}(T)=|\Psi (T)\rangle\langle\Psi (T)|$.
Nevertheless, the information on the atomic system solely  can be
obtained from the atomic reduced density matrix $\hat{\rho}_a(T)$
having the form

\begin{eqnarray}
\begin{array}{lr}
\hat{\rho}_a(T)={\rm Tr}_f\hat{\rho}(T),\\
\\
\hat{\rho}_{a}(T) =\hat{\rho}_{ee}(T)|e\rangle\langle
e|+\hat{\rho}_{gg}(T)|g\rangle\langle g|+
\hat{\rho}_{eg}(T)|e\rangle\langle
g|+\hat{\rho}_{eg}^{*}(T)|g\rangle\langle e|,
 \label{l6}
 \end{array}
\end{eqnarray}
where the subscript $f$ means that we trace out the field and
$\hat{\rho}_{ij}(T)=\langle i|\hat{\rho}_a(T)|j\rangle$ with
$i,j=e,g$. Form (\ref{a8a}) the coefficients $\hat{\rho}_{ij}(T)$
can be expressed as:
\begin{eqnarray}
\begin{array}{lr}
\rho_{ee}(T)=\sum\limits_ {n=0}^{\infty}
|C_n|^2\cos^2(T\sqrt{n+1}),\quad \rho_{gg}(T)=\sum\limits_
{n=0}^{\infty} |C_n|^2\sin^2(T\sqrt{n+1}),\\
\\
\rho_{eg}(T)=i\exp(i\vartheta)\sum\limits_ {n=0}^{\infty}
|C_{n+1}C_n|\cos(T\sqrt{n+2})\sin(T\sqrt{n+1}).
 \label{ll6}
 \end{array}
\end{eqnarray}
Additionally, for the atomic set operators $\{\hat{\sigma}_x,
\hat{\sigma}_y,
 \hat{\sigma}_z\}$ we obtain:
\begin{eqnarray}
\begin{array}{lr}
\langle\hat{\sigma}_{z}(T)\rangle=\rho_{ee}(T)-\rho_{gg}(T),\quad
 \langle
\hat{\sigma}_{x}(T)\rangle=2{\rm Re}[\rho_{eg}(T)],\\
\\
 \langle \hat{\sigma}_{y}(T)\rangle=2{\rm Im}[\rho_{eg}(T)],
\qquad \rho_{ee}(T)+\rho_{gg}(T)=1.
 \label{ll6y}
 \end{array}
\end{eqnarray}

Now the linear entropy is defined as:
\begin{equation}
\xi(T)=1-{\rm Tr}\rho^2_a(T).
 \label{law1}
\end{equation}
 For the system under consideration the relation
(\ref{law1}) by means of (\ref{ll6y}) can be easily evaluated as:
\begin{eqnarray}
\begin{array}{lr}
\xi(T)=1-
\rho^2_{ee}(T)-\rho^2_{gg}(T) -2|\rho_{eg}(T)|^2\\
\\
= \frac{1}{2}[1-\eta^2(T)], \qquad
\eta^2(T)=\langle\hat{\sigma}_x(T)\rangle^2+\langle\hat{\sigma}_y(T)\rangle^2+
\langle\hat{\sigma}_z(T)\rangle^2,
 \label{laww2}
 \end{array}
\end{eqnarray}
where the quantity $\eta(T)$ is  the Bloch sphere radius. The Bloch
sphere is a well-known tool in quantum optics, where the simple
qubit state is faithfully represented, up to an overall phase, by a
point on a standard sphere with radius unity, whose coordinates are
expectation values of the atomic set operators of the system.
 In the language of entanglement $\xi(T)$
ranges from $0$ ( i.e., $\eta(T)=1$) for disentangled and/or pure
states to $0.5$ ( i.e., $\eta(T)=0$) for maximally entangled
bipartite \cite{julio}. On the other hand, the von Neumann entropy
is defined as \cite{neum}:
\begin{eqnarray}
\begin{array}{lr}
\gamma(T)=-{\rm Tr}\{\rho_a(T){\rm ln}\rho_a(T)\},\\
\\
=-\mu_{-}(T){\rm ln}\mu_{-}(T)-\mu_{+}(T){\rm ln}\mu_{+}(T),
 \label{law3}
 \end{array}
\end{eqnarray}
where $\mu_{\pm}(T)$ are the eigenvalues of the $\rho_a(T)$, which
for (\ref{6}) can be expressed as:
\begin{eqnarray}
\begin{array}{lr}
\mu_{\pm}(T)=\frac{1}{2}\{1\pm\sqrt{1-4[\rho_{ee}(T)\rho_{gg}(T)
-|\rho_{eg}(T)|^2]}\},\\
\\ = \frac{1}{2}\{1\pm\eta(T)\}.
 \label{law2}
\end{array}
\end{eqnarray}
The second  line in (\ref{law2}) has been evaluated by means of
(\ref{ll6}) and (\ref{ll6y}). From the limitations on the $\eta(T)$
one can prove $0\leq\gamma(T) \leq 0.693$. Finally, the atomic Wehrl
entropy  has been defined as \cite{karol}:
\begin{equation}
W_{a}(T)=-\int_{0}^{2\pi }\int_{0}^{\pi } Q_{a}(\Theta ,\Phi ,T)\ln
Q_{a}(\Theta ,\Phi ,T)\sin \Theta d\Theta d\Phi, \label{aw8}
\end{equation}%
where $Q_{a}(\Theta ,\Phi ,T)$ is
 the atomic $Q$-function defined as:

\begin{equation}
Q_{a}(\Theta ,\Phi ,T)=\frac{1}{2\pi }\left\langle \Theta ,\Phi
\left\vert \hat{\rho}_{a}(T)\right\vert \Theta ,\Phi \right\rangle
\label{aw1}
\end{equation}%
and
 $\left\vert \Theta ,\Phi \right\rangle $ is the
atomic coherent state having the form \cite{vie}:
\begin{equation}
\left\vert \Theta ,\Phi \right\rangle =\cos \left( \Theta /2\right)
\left\vert e \right\rangle +\sin \left( \Theta /2\right) \exp (i\Phi
)\left\vert g \right\rangle \label{aw3}
\end{equation}
with $0\leq \Theta\leq \pi, 0\leq \Phi \leq 2\pi$. For the
wavefunction (\ref{a8a}) the atomic $Q_a$ function can be
evaluated
 as
\begin{eqnarray}
\begin{array}{lr}
Q_{a}(\Theta ,\Phi ,T)=\frac{1}{4\pi }[1+\beta(T)],\\
\\
\beta(T) =\langle\hat{\sigma}_{z}(T)\rangle\cos \Theta +\left[
\langle\hat{\sigma}_{x}(T)\rangle\cos \Phi
+\langle\hat{\sigma}_{y}(T)\rangle\sin \Phi \right] \sin \Theta.
\label{aw4}
\end{array}
\end{eqnarray}
One can easily check that  $Q_a$ given by (\ref{aw4}) is normalized.
 On substituting (\ref{aw4}) into (\ref{aw8}) and
carrying  out the integration we obtain

\begin{eqnarray}
\begin{array}{lr}
W_{a}(T)={\rm ln}(4\pi)-\frac{1}{4\pi } \int_{0}^{2\pi
}\int_{0}^{\pi }[1+\beta(T)]{\rm ln}[1+\beta(T)] \sin \Theta d
\Phi d \Theta,\\
\\
={\rm
ln}(4\pi)-\sum\limits_{n=1}^{\infty}\sum\limits_{r=0}^{n}\sum\limits_{s=0}^{r}
\frac{(2n)!(-1)^s\langle\hat{\sigma}_{z}(T)\rangle^{2(n-r)}[\langle\hat{\sigma}_{x}(T)\rangle^2
 +\langle\hat{\sigma}_{y}(T)\rangle^2]^r
}{2n(2n-1)(2n-2r)!r!4^r (r-s)!s![2(n+s-r)+1]}. \label{alw4}
\end{array}
\end{eqnarray}
In the derivation of  (\ref{alw4}) we have used the series expansion
of the logarithmic function, the binomial expansion  and the
identity \cite{tabl}:

\begin{equation}
\int_{0}^{2\pi}(c_1\sin x +c_2\cos x)^kdx= \left\{
\begin{array}{lr}
0
\;\;&{\rm for}\;k=2m+1 ,\\
2\pi\frac{(2m)!}{4^m (m!)^2} (c_1^2+c_2^2)^m
 \;\;&{\rm
for}\;k=2m,
\end{array}
\right. \label{8a}
\end{equation}
where $c_1, c_2$ are c-numbers and $k$ is positive integer.
Expression (\ref{alw4}) is relevant for the numerical
investigation. In Figs. 1(a), (b) and (c) we have plotted the
linear entropy (\ref{laww2}),
 the von Neumann entropy (\ref{law3}) and the atomic Wehrl entropy
(\ref{alw4}), respectively, for the given values of the interaction
parameters.
\begin{figure}
    \includegraphics[width=1.0\linewidth]{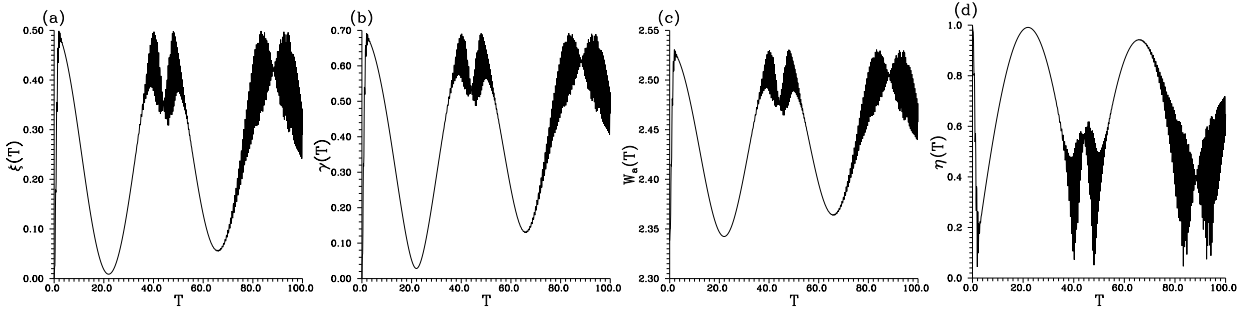}
\caption{
 The linear entropy (a), the von Neumann entropy
 (b),  the atomic Wehrl entropy (c) and the Bloch sphere radius $\eta(T)$ (d)  for
  $(|\alpha|,\phi)=(7,0)$. }
\end{figure}
Comparison between these figures is instructive and shows that the
three entropic relations provide typical information (with
different scales) on the entanglement in the JCM. Now we'll
explain why this occurs. We start by expressing (\ref{law3}) in a
series form using the Taylor expansion for the logarithmic
functions, where we obtain
\begin{equation}
\gamma(T)={\rm ln}2-\sum\limits_{n=1}^{\infty
}\frac{\eta^{2n}(T)}{2n(2n-1)}. \label{awl8}
\end{equation}%
  From
(\ref{laww2}) and (\ref{awl8}) one can realize that $\xi(T)$ and
$\gamma(T)$ are functions in  $\eta^2(T)$ and hence they exhibit
similar behaviors.
 Now we draw the attention to (\ref{alw4}), which
is rather complicated. Nevertheless, by expanding the first few
terms in this expression one can obtain:

\begin{eqnarray}
\begin{array}{lr}
W_{a}(T) ={\rm ln}(4\pi)-\{
\frac{\eta^2(T)}{1\times 2\times 3}+\frac{\eta^4(T)}{3\times 4\times5}+
\frac{\eta^6(T)}{5\times 6\times 7}+...\}\\
\\
={\rm ln}(4\pi)-
\sum\limits_{n=1}^{\infty}\frac{\eta^{2n}(T)}{2n(2n-1)(2n+1)}.
\label{alw4w}
\end{array}
\end{eqnarray}
 The series in the second line  of (\ref{alw4w}) has been
obtained from the first one via mathematical induction.  From
(\ref{alw4w}) it is obvious that $W_a(T)$ is a function in
$\eta^2(T)$ and this is the reason for having behavior  similar to
those of $\xi(T)$ and $\gamma(T)$. Also from the available
information on $\eta(T)$ and  (\ref{alw4w}) one can explore the
limitations of $W_a(T)$. For instance,  for maximal entangled
bipartite we have $\eta(T)=0$ and hence the upper bound of the
$W_a(T)$ is ${\rm ln}(4\pi)$, however, for disentangled bipartite
(, i.e. $\eta(T)=1$)  the lower bound can be evaluated as:
\begin{equation}
W_a(T)={\rm
\ln}(4\pi)-\sum\limits_{n=1}^{\infty}\frac{1}{2n(2n-1)(2n+1)}\simeq
{\rm \ln}(4\pi)-0.19315\simeq 2.3379. \label{lim1}
\end{equation}
The exact value of the series in (\ref{lim1})  has been evaluated
numerically.  All  the analytical facts obtained above are
remarkable in Fig. 1(c).

We conclude this Letter by deriving a closed form for
(\ref{alw4w}). By means of the partial fraction one can prove
obtain
\begin{equation}
\frac{1}{2n(2n-1)(2n+1)}=\frac{1}{2n(2n-1)}-\frac{1}{2(2n-1)}+\frac{1}{2(2n+1)}.
\label{par}
\end{equation}
Substituting (\ref{par}) into (\ref{alw4w}) and using the
identities:
\begin{equation}
\sum\limits_{n=1}^{\infty}\frac{\eta^{2n}(T)}{2n(2n-1)}=
\frac{1}{2}{\rm ln}[1-\eta^2(T)]+ \frac{\eta(T)}{2}{\rm
ln}\left[\frac{1+\eta(T)}{1-\eta(T)}\right], \label{par1}
\end{equation}

\begin{equation}
\sum\limits_{n=1}^{\infty}\frac{\eta^{2n}(T)}{(2n-1)}=\frac{\eta(T)}{2}
{\rm ln}\left[\frac{1+\eta(T)}{1-\eta(T)}\right], \quad
\sum\limits_{n=1}^{\infty}\frac{\eta^{2n}(T)}{(2n+1)}=-1+\frac{1}{2\eta(T)}
{\rm ln}\left[\frac{1+\eta(T)}{1-\eta(T)}\right],  \label{par2}
\end{equation}
we obtain
\begin{equation}
W_{a}(T) =\frac{1}{2}+{\rm ln}(4\pi)- \frac{1}{2}{\rm
ln}(1-\eta^2(T))+
\frac{1}{4}\left[\eta(T)+\frac{1}{\eta(T)}\right]{\rm
ln}\left[\frac{1-\eta(T)}{1+\eta(T)}\right]. \label{faal}
\end{equation}
The identity (\ref{par1}) is obtained from (\ref{law3}) and
(\ref{awl8}), however, those in (\ref{par2}) are related to the
logarithmic form of the $\tanh^{-1}(.)$. The expression
(\ref{faal}) is valid for all values of  $\eta(T)$ except
$\eta(T)=0$. We have checked the validity  of the expression
(\ref{faal}) by getting the typical behavior  shown in Fig. 1(c).

Now, the different scales in the above entropic relations can be
treated by redefining  them as follows:
\begin{equation}\label{yous}
\widetilde{\gamma}(T)=\frac{\gamma(T)}{0.693},\qquad
\widetilde{W}_{a}(T)=\frac{{\rm ln}(4\pi)-W_{a}(T)}{0.19315}.
\end{equation}
In this case, the enropic relations $\xi(T),
\widetilde{\gamma}(T)$ and $\widetilde{W}_{a}(T)$ yield typical
information on the JCM. As all forms of
  the two-level JCM (, i.e. off-resonance JCM, multimode JCM, etc.)
  can be described by the atomic density matrix
 (\ref{l6}), the results obtained in this Letter are universal.
 From the above investigation
 the quantities $\xi(T), \gamma(T)$ and  $ W_{a}(T)$ depend only on
  $\eta(T)$ and hence it would be more convenient to use the Bloch
  sphere radius $\eta(T)$ directly for getting information on the
  entanglement in the JCM (see Fig. 1(d)). In this case, for maximum (minimum) entanglement
    we have $\eta(T)=0\quad (1)$, i.e., $0\leq \eta(T)\leq 1$. It is
    worth mentioning that the concept of the Bloch sphere has been used recently
    for the JCM with different types of initial states for analyzing the
    correlation between entropy exchange and entanglement \cite{bouk}.
The final remark,  the linear entropy, the von Neumann entropy and
the atomic Wehrl operator of the JCM are invariant quantities
under unitary transformations, as they depend only on the
eigenvalues of the density operator.

  In conclusion in this Letter we have  analytically explained why the linear
entropy, the von Neumann entropy and the atomic Wehrl entropy of
the JCM have similar behaviors. We have shown that the Bloch
sphere radius has to be used for quantifying the entanglement in
the JCM instead of the entropic relations. Finally, the results
obtained in this Letter are universal.

\section*{ Acknowledgement}

 The author would like to thank the Abdus Salam International
Centre for Theoretical Physiscs, Strada Costiers, 11 34014 Trieste
Italy for the hospitality and financial support under the system of
associateship, where a part of this work is done.

\end{document}